\documentstyle[aps,preprint,epsf]{revtex}

\def\ltap{\ \raise.3ex\hbox{$<$\kern-.75em\lower1ex\hbox{$\sim$}}\ }
\def\gtap{\ \raise.3ex\hbox{$>$\kern-.75em\lower1ex\hbox{$\sim$}}\ }

\def\CM{{\cal M}}
\def\inbar{\vrule height1.5ex width.4pt depth0pt}
\def\IP{\relax{\rm I\kern-.18em P}}
\def\IR{\relax{\rm I\kern-.18em R}}
\def\IC{\relax\hbox{\kern.25em$\inbar\kern-.3em{\rm C}$}}

\newcommand{\bel}[1]{\begin{equation}\label{#1}}
\newcommand\ee{\end{equation}}
\newcommand{\eref}[1]{(\ref{#1})}
\newcommand{\Eref}[1]{Eq.~(\ref{#1})}
\newcommand{\SEC}[1]{Sec.~\ref{sec:#1}}
\newcommand{\SSEC}[1]{Sec.~\ref{ssec:#1}}

\def\rarr{\rightarrow}
\def\none{$\CN=1$}

\def\susy{supersymmetry}

\def\susic{supersymmetric}

\newcommand{\rem}[1]{}

\def\be{\begin{equation}}
\def\ee{\end{equation}}

\def\half{{1\over 2}}
\def\NN{{\cal N}}

\def\none{$\NN=1$}
\def\ntwo{$\NN=2$}

\def\nfour{$\NN=4$}
\def\susy{supersymmetry}
\def\susic{supersymmetric}

\def\rarr{\rightarrow}

\def\ZZ{{\bf Z}}

\def\OM{Olive-Montonen}

\def\SL{$SL(2,\ZZ)$}

\newcommand{\drawsquare}[2]{\hbox{%
\rule{#2pt}{#1pt}\hskip-#2pt
\rule{#1pt}{#2pt}\hskip-#1pt
\rule[#1pt]{#1pt}{#2pt}}\rule[#1pt]{#2pt}{#2pt}\hskip-#2pt
\rule{#2pt}{#1pt}}

\newcommand{\fund}{\raisebox{-.5pt}{\drawsquare{6.5}{0.4}}}
\newcommand{\antifund}{\overline{\fund}}

\begin{document}
\pagestyle{empty}

\preprint{
\begin{minipage}[t]{3in}
\begin{flushright}
{IASSNS--HEP--98/23 \\
hep-ph/9803086 \\ \ \\
}
\end{flushright}
\end{minipage}
}

\title{Finite Theories and Marginal Operators on the Brane}

\author{Amihay Hanany, Matthew J. Strassler, Angel M. Uranga\\ 
{\tt hanany, strassler, uranga@ias.edu} }

\address{School of Natural Sciences\\
Institute for Advanced Study\\
 Princeton, NJ 08540, USA}

\maketitle
\begin{abstract}
We show how to use D and NS fivebranes in Type IIB superstring theory
to construct large classes of finite \none\ supersymmetric four dimensional
field theories.  In this construction, the beta functions of the
theories are directly related to the bending of branes; in finite
theories the branes are not bent, and vice versa.  Many of these
theories have multiple dimensionless couplings.  A group of duality
transformations acts on the space of dimensionless couplings; for a
large subclass of models, this group always includes an overall
$SL(2,\ZZ)$ invariance. In addition, we find even larger classes of
theories which, although not finite, also have one or more marginal
operators.
\end{abstract}
\pacs{PACS}

\tightenlines

Most quantum field theories do not contain adjustable
dimensionless couplings.  Even conformal field theories are generally
isolated points in the space of all possible models.  Field theories
with exactly marginal operators, including finite theories, are
interesting because they contain truly dimensionless continuous
parameters.  Many examples are known in two, three and four
dimensions.  Our focus in this paper will be the four-dimensional case.

One reason why dimensionless couplings are worthy of study is that
duality transformations can act on them.  The classic example is the
action of $SL(2,\ZZ)$ on the coupling $\tau = {\theta\over
2\pi}+i{4\pi\over g^2}$ in \nfour\ \susic\ theories.  \ntwo\ and
\none\ \susic\ gauge theories are known in which a duality group $G_D$
acts on a complex coupling constant
\cite{nsewtwo,sundual,sundualb,witMa,emop,kinstwo,aks,asty,ails}.  In
the \none\ case, many of the theories with marginal couplings are not
finite; instead, these theories have non-trivial beta functions whose
zeroes occur when some fields' anomalous dimensions are non-zero
\cite{emop,kinstwo,aks}.  Often finite theories with marginal
couplings are dual (using \none\ duality \cite{NAD}) to non-finite
theories with marginal couplings \cite{emop,kinstwo,karchlust}.  If
one wants to characterize the space of theories and how different
duality transformations act on this space --- which may be relevant
for the recent work relating string theory in ten dimensions to gauge
theory in four --- then understanding these theories is important.
\footnote{ It has also long been argued that finite grand unified
theories might be relevant for phenomenology; for a recent discussion
see \cite{finiteGUT}.  However, given the existence of gravity as a
natural cutoff for field theory, and given the finiteness of string
theory, we see no particular compelling argument for finite field
theories below the string scale.}

Recently, constructions of field theories using branes of string
theory and M theory have proven to be a powerful tool for studying
properties both of the field theories and of the branes; for a recent
review, see \cite{gkreview}.  In this paper, we show that the Type IIB
fivebrane constructions of \none\ \susic\ field theories can be used
to generate large classes of finite models, and even larger classes of
models which are not finite but still have marginal operators.  In
\SEC{fieldtheory} we discuss the criteria for marginal couplings in
general and finiteness in particular.  Next, in \SEC{branes}, we
discuss the brane construction of a large set of field theories.  We
then identify a class of theories in \SEC{margops} which satisfy the
criteria for containing at least one marginal operator, and consider a
number of examples, many of which are finite.  We will see both
``elliptic'' models (generalizations of \nfour\ supersymmetric $SU(N)$
gauge theories) and ``cylindrical'' models (generalizations of \ntwo\
supersymmetric $SU(N)$ theories with $2N$ hypermultiplets) which are
degenerations of the elliptic models.  We discuss the duality
properties of these models in \SEC{duality}.  As yet we have only a
partial understanding of the coupling constants and duality
transformations that these theories possess; we discuss some
unresolved issues in Appendix A.  Proofs of various claims appear in
Appendix B.\footnote{After this work was substantially completed, a
number of papers on related subjects appeared \cite{Ibanez,ks,lnv}
whose motivation is completely different but whose results are related
to our own.  The finite models discussed here appeared in the
classification of \cite{lnv}.  More closely related work appeared in
another recent paper \cite{gimgrem}.}

\section{Field Theory Considerations}
\label{sec:fieldtheory}

 In four dimensions all known examples of theories with marginal
operators have four or more supersymmetries.\footnote{Proposals for
finite non-supersymmetric field theories have recently been made
\cite{ks,lnv}, though they likely are finite only when the number of
colors is strictly infinite.}  Finite theories in \none\ \susy\ have a
long history and proofs have been given that they are finite to all
orders.  Particularly clear are the proofs of Lucchesi, Piguet and
Sibold \cite{LPSa,LPSb}.  However, the approach of \cite{emop}, which
is similar in some ways to \cite{LPSa,LPSb}, has the advantages that
it is extremely simple to state, is non-perturbative in form, and
generalizes to non-finite theories.

There is a very simple criterion that tests whether an \none\ \susic\
theory may have a marginal operator \cite{emop}.  More details
and additional references are given in \cite{emop,matrev}.  If
the superpotential contains a term $W=h\phi_1\dots\phi_n$, then the
beta function corresponding to the coupling $h$ is
\bel{betah}
\beta_h
\propto A(h) \equiv (n-3)+\half\sum_{k=1}^n
\gamma(\phi_k)
\ee
where $\gamma(\phi_k)$ is the anomalous mass dimension of the
superfield $\phi_k$.  For the gauge coupling of a gauge group $G$, the
formula for the beta function is \cite{NSVZ,SVa,SVb}
\bel{betaSV}
\beta_g \propto A(g) \equiv -\left\{ \left[3C_2(G)- \sum_j T(R_j)\right]
                   +\sum_j T(R_j) \gamma(\phi_j) \right\}
\ee
where the sums are over all matter fields, $C_2(G)$ is the quadratic
Casimir of the adjoint representation of $G$, and $T(R_j)$ is the
quadratic Casimir of the representation $R_j$ of $G$ in which $\phi_j$
transforms.  The term in square brackets is the one-loop beta function
coefficient $b_0$.

 If we have a theory with $p$ Yukawa couplings $h_i$ and $q$ gauge
couplings $g_j$, the anomalous dimensions are unknown real functions
of the couplings $h_i$ and $g_j$.  The condition for a conformal fixed
point is that all beta functions $\beta_{h_i}$, $\beta_{g_j}$ vanish,
putting $p+q$ conditions on $p+q$ couplings.  The generic theory has
isolated solutions to these conditions.  However, because {\it the
beta functions are linear functionals of the anomalous dimensions}, it
may happen that the $p+q$ conditions $\beta_{h_i}= 0=\beta_{g_j}$ are
linearly dependent.  If only $p+q-r$ of these conditions are linearly
independent, then the solutions will form an $r$ complex-dimensional
subspace of the $p+q$ complex dimensional space of couplings.  These
solutions will represent a set of conformal field theories with $r$
adjustable dimensionless couplings.  Of course, a given theory may
have {\it no} interacting conformal fixed points anywhere in coupling
space.

If the conditions for a fixed point also imply that all the anomalous
dimensions vanish, then the theory has no divergences in perturbation
theory (except those which appear in composite operators) and is as
finite as \nfour\ \susic\ gauge theory.  In this case the manifold of
fixed points necessarily includes the free theory.  Note that two
necessary conditions for finiteness are that all couplings in the
superpotential be dimensionless at the classical level and that all
gauge beta functions vanish at one loop.\footnote{It is proven that if
there is a choice of couplings such that all beta functions vanish and
all anomalous dimensions vanish at leading order, then the theory is
finite to all orders \cite{LPSb}.  There is also a proof \cite{matunp}
that if the manifold of fixed points includes the free theory, then
the anomalous dimensions vanish at leading order.  Thus, when the
conditions of \cite{emop} for vanishing beta functions permit the
possibility that the anomalous dimensions vanish, that possibility is
apparently always realized.  A profound understanding of why this is
true is lacking.}

Lists of finite models may be found in \cite{jzlist}.  Many examples of
theories with marginal operators, including finite models, are given
in \cite{emop}.  Here we mention a couple of well-known cases.
Consider $SU(3)$ with nine triplets $Q^i$ and nine antitriplets
$\tilde Q_j$ \cite{fsua,fsub,fsuc}. This theory is finite if the
superpotential
\bel{suthree}
 W=h\left(
Q^{1}Q^{2}Q^{3}+Q^{4}Q^{5}Q^{6}+Q^{7}Q^{8}Q^{9}+ \tilde Q_{1}\tilde
Q_{2}\tilde Q_{3}+\tilde Q_{4}\tilde Q_{5}\tilde Q_{6}+\tilde Q_{7}\tilde
Q_{8}\tilde Q_{9}\right)
\ee
is added.  The flavor symmetries among the matter fields ensure they
all have the same anomalous dimension $\gamma(g,h)$. (Note that
if the flavor symmetries were slightly broken by the superpotential,
they would be restored in the infrared \cite{emop}.) Since
$\beta_g\propto \beta_h\propto\gamma(g,h)$, the theory is conformal
and has no perturbative divergences when $\gamma(g,h) =0$ --- a single
condition on two couplings, leading to a one-complex-dimensional
space of solutions.  It is easy to show that $\gamma$ has a zero in
perturbation theory.

Other similar examples include $E_6$ with twelve fields $Q^i$
in the  ${\bf 27}$ representation with superpotential
\be
W=h_1 \sum_{r=1}^{12} (Q^r)^3 +
h_2\sum_{r=1}^{4}Q^{r}Q^{r+1}Q^{r+2}
\ee
and $SU(N)$ theories
with \nfour\ matter content (three adjoint chiral
multiplets $\Phi_1,\Phi_2,\Phi_3$)  with superpotential
\be
W = (h_1\ f^{abc}+h_2\ d^{abc})\Phi_1^a\Phi_2^b\Phi_3^c
+ h_3d^{abc} \sum_{i=1}^3 \Phi^a_i\Phi^b_i\Phi^c_i
\ee
Both of these have multiple marginal operators.  Note
also that the $E_6$ model is chiral.

An example of a model that is not finite but has a marginal
operator is $SU(4)$ with eight flavors and a quartic
superpotential \cite{emop}.
\bel{sufourB}
W= h(Q^{1}Q^{2}Q^{3}Q^{4}+Q^{5}Q^{6}Q^{7}Q^{8}+
      \tilde Q_{1}\tilde Q_{2}\tilde Q_{3}\tilde Q_{4}
+\tilde Q_{5}\tilde Q_{6}\tilde Q_{7}\tilde Q_{8}).
\ee
This model is a continuous deformation of the theory with vanishing
superpotential, which is believed to have a fixed point at some value
of the gauge coupling \cite{NAD}.

There is also a link between marginal operators and the duality of a
wide class of \none\ models. In \cite{emop}, a close connection was
noted between the duality of finite \ntwo\ $SU(N)$ theories
\cite{nsewtwo,sundual,sundualb} and duality of non-finite \none\
$SU(N)$ models with marginal operators.  This connection was further
explored in \cite{kutsch,lssosp,ils,brostra}.  

We now turn to the construction of large classes of these models, both
finite and otherwise, using fivebranes in Type IIB string theory. It
turns out that the rules for finiteness and for marginal operators in
these theories translate into simple geometrical and algebraic
statements.

\section {Brane Construction}
\label{sec:branes}

\subsection{Field theories from fivebranes}
\label{ssec:fieldfive}

The finite \none\ models we will be considering are constructed in a
brane setup, in the spirit of \cite{hw}, which was described in detail
in \cite{hz}.  The description here will be short and further details
are contained in \cite{hz}.

We are working in Type IIB superstring theory with the following set
of branes.
\begin{itemize}
\item NS branes along 012345 directions
\item NS$'$ branes along 012367 directions
\item D5 branes along 012346 direction.
\end{itemize}
The D5 branes will be finite in two of the directions, 4 and 6; their
low-energy effective world volume theory is 3+1 dimensional.  The
presence of all branes breaks supersymmetry to 1/8 of the original
supersymmetry, and thus we are dealing with \none\ supersymmetry (4
supercharges) in four dimensions.  

In most of the applications in this paper we will be working on a
torus in the 46 directions.  We consider the 46 plane $\IR^2$ divided
into a grid by an infinite number of NS and NS$'$ branes. In each box
of the grid we will place a number of D5 branes.  If the assemblage
has translational symmetries, which form a lattice group $\Lambda$
generated by two shift vectors $v_1$ and $v_2$, then we can construct
a torus $T^2= \IR^2/\Lambda$ which consists of the unit 
cell of the lattice $\Lambda$.   The 4 and 6 directions will become
circles with radii $R_4$ and $R_6$.  

\begin{figure}
\centering
\epsfxsize=3.5in
\hspace*{0in}\vspace*{.2in}
\epsffile{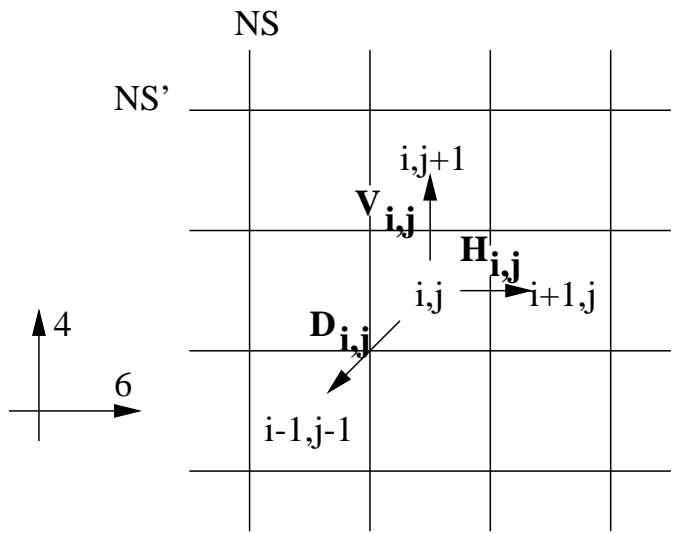}
\caption{The NS and NS$'$ branes form a grid; in each box $(i,j)$ of
the grid lie $n_{i,j}$ D5 branes.  The arrows denote the chiral
multiplets $H_{i,j},V_{i,j},D_{i,j}$ which are in the fundamental of
the group $SU(n_{i,j})$ and in the antifundamental of an adjacent
group.  The arrows at the far left indicate the $x^4$ and $x^6$
coordinates.}
\label{fig:multiplets}
\end{figure}

A generic configuration then consists of a grid of $k$ NS branes and $k'$
NS$'$ branes, dividing the 46 torus into a set of
$k k'$ boxes. In each box, we can place an arbitrary number of D5
branes. Let $n_{i,j}$ denote the number of D5 branes in the box $i,j$,
$i=1, \ldots, k, j=1, \ldots, k'$.  In the following, indices will
denote variables in a periodic fashion: an index $i$ is to be
understood modulo $k$ and an index $j$ is to be understood modulo
$k'$.  Thus a model's gauge and matter content is specified by the
numbers $k$ and $k'$ and the set of numbers $\{n_{i,j}\}$.

The gauge group is $\prod_{i,j}SU(n_{i,j})$.  (Classically the gauge
group also includes one $U(1)$ factor for each box, though these are
not present quantum mechanically; see the discussion in \SSEC{FIterms}.)
The matter content of the model consists of three types of \none\ chiral
representations.  They will be denoted as $H_{i,j}$, $V_{i,j}$ and
$D_{i,j}$, corresponding to the horizontal, vertical and diagonal
multiplets which arise in the brane system (see the details in
\cite{hz}). $H_{i,j}$ transforms in the $(\fund,\antifund)$ of
$SU(n_{i,j})\times SU(n_{i+1,j})$, $V_{i,j}$ transforms in the
$(\fund,\antifund)$ of $SU(n_{i,j})\times SU(n_{i,j+1})$ and $D_{i,j}$
transforms in the $(\fund,\antifund)$ of $SU(n_{i,j})\times
SU(n_{i-1,j-1})$. Figure \ref{fig:multiplets} shows the conventions
for denoting the multiplets.

\begin{figure}
\centering
\epsfxsize=3in
\hspace*{0in}\vspace*{.2in}
\epsffile{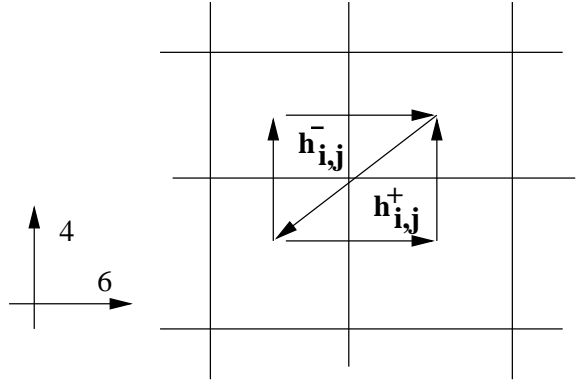}
\caption{The two superpotential interactions at each corner, with
couplings $h^+_{i,j}$ and $h^-_{i,j}$, are represented by an oriented
triangle of arrows.}
\label{fig:superpot}
\end{figure}

The superpotential in these models is calculated using the rules
described in \cite{hz}.  It is given by
\bel{HVD}
W=\sum_{i,j}H_{i,j}V_{i+1,j}D_{i+1,j+1}
-\sum_{i,j}H_{i,j+1}V_{i,j}D_{i+1,j+1}.
\ee
The first term corresponds to lower triangles of arrows and the second
term corresponds to upper triangles of arrows in the notation of
\cite{hz}, as shown in figure \ref{fig:superpot}.  Note the relative
minus sign between the two terms. By symmetry, the signs are obviously
independent of $i$ and $j$; the relative sign is determined by looking
at the special case $k=k'=1$, which will be discussed in more detail
below.  From a general field theory point of view, there may be
coefficients $h^+_{i,j}$ and $h^-_{i,j}$ for each Yukawa term.  It is
convenient for now to set these coefficients to one by a redefinition
of the fields.

Our understanding of the gauge couplings, especially the theta angles,
is incomplete.  However, some contributions to these couplings can be
characterized.  In the simplest situations, the gauge couplings of the
various gauge groups are given by the positions of the NS branes in
the $x^6$ direction and the position of the NS$'$ branes in the $x^4$
direction.  There are $k$ positions $x_6^i$ and $k'$ positions
$x_4^j$.  Correspondingly, the $x_6$ direction is divided into $k$
intervals with lengths $a_i=x_6^i-x_6^{i-1}$, such that
$\sum_ia_i=R_6$.  The $x_4$ direction is divided into $k'$ intervals
of length $b_j=x_4^j-x_4^{j-1}$, such that $\sum_jb_j=R_4$.  The gauge
coupling $g_{i,j}$ for the group $SU(n_{i,j})$ is given by
\bel{gauge}
{1\over g_{i,j}^2} = {a_i b_j \over g_s l_s^2}.
\ee
The $kk'$ gauge couplings are not all independent. They are given by
$k+k'-1$ parameters corresponding to the positions of the NS and NS$'$
branes.  Two positions can be set to zero by the choice of origin in
the 46 directions, but the area of the 46 torus gives one more
parameter. The couplings do not depend on the ratio between the two
radii of the torus.  As we will see later, the field theories often
have more than $k+k'-1$ dimensionless parameters, indicating that we have not
identified all of the contributions to these couplings.

The theta angles of the gauge theories receive various
contributions. Let $A_i$ be the gauge field on the world volume of the
$i^{th}$ NS brane and $A'_j$ be the gauge field on the world volume of
the $j^{th}$ NS$'$ brane. Since the dimensions 4 and 6 are compact,
there can be non-zero Wilson lines of $A_i$ along 4, and of $A'_j$
along 6.  Let $R_{i,j}$ denote the area in the 46 direction which is
bounded by the pair of NS branes and NS$'$ branes. The theta angle for
the $i,j$ group depends on the line integral of the different gauge
fields along the boundary of $R_{i,j}$. Schematically,
\bel{theta}
\theta_{i,j}= \int_{R_{i,j}}B + \int_{a_i}(A'_{j-1}-A'_{j}) +
\int_{b_{j}}(A_{i}-A_{i-1}).
\ee
where $B$ is the RR two form of Type IIB superstring theory. The
contributions from the gauge fields are required for the invariance of
$\theta_{i,j}$ under gauge transformations of $B$.  Were this the
entire story we would again have $k+k'-1$ parameters.  However,
invariance under gauge transformations of the one-forms require that
additional terms be added to this expression involving axion-like
fields living at the intersections of the NS and NS$'$ branes.  Some
additional discussion of this issue is presented in Appendix A.

\subsection{An example}

\begin{figure}
\centering
\epsfxsize=3in
\hspace*{0in}\vspace*{.2in}
\epsffile{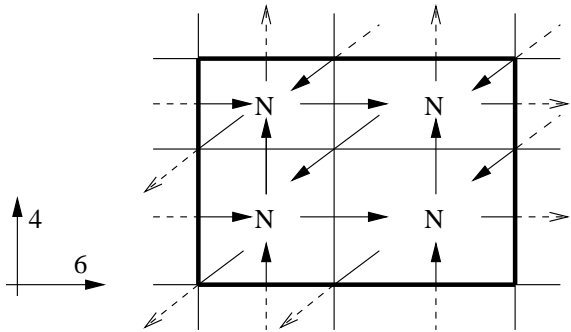}
\caption{Construction of a model on a torus, expressed
as a unit cell of a two-dimensional lattice.  The unit cell of four boxes is
highlighted.  Note the twelve matter
multiplets, indicated by the solid parts of the arrows; the dotted
parts of the arrows correspond to their images under shifts of the
unit cell by lattice vectors.}
\label{fig:twobytwo}
\end{figure}

Let us specify this in a concrete case, with two NS branes of each
kind, $k=k'=2$, and an equal number $N$ of D5 branes in each box.  The
torus is the grid identified under shifts by two boxes vertically and
by two boxes horizontally.  The unit cell's four boxes, eight Yukawa
interactions and twelve matter fields are shown in figure
\ref{fig:twobytwo}.  This gives an \none\ gauge theory with gauge
group $SU(N)\times SU(N)\times SU(N)\times SU(N)$, with vectorlike
matter content
\bel{fourboxmatter} \begin{array}{ccccc}
H_{1,1} & = & \fund,\antifund,1,1 & = \overline{H_{2,1}} \\
V_{1,1} & = & \fund,1,\antifund,1 & = \overline{V_{1,2}} \\
D_{1,1} & = & \fund,1,1,\antifund & = \overline{D_{2,2}} \\
H_{1,2} & = & 1,1,\fund,\antifund & = \overline{H_{2,2}} \\
V_{2,1} & = & 1,\fund,1,\antifund & = \overline{V_{2,2}} \\
D_{2,1} & = & 1,\fund,\antifund,1 & = \overline{D_{1,2}} \\
\end{array}
\ee
The superpotential is as in \Eref{HVD}.  In the simplest situation, the
four gauge couplings are specified by the areas of the boxes.

\subsection{Bending and the Beta Function}
\label{ssec:bending}

It is important to note that there is a geometrical picture for the
beta function in this class of models.  This is a well-known property
of theories with eight supercharges, which we review.  Consider the
constructions of field theories using D branes on a one-dimensional
interval (as opposed to the two-dimensional intervals used in this
paper), as in \cite{hw} and its various generalizations; see
\cite{gkreview} for a review.  A field theory in $d$ dimensions with 8
supercharges is realized on the world volume of a Dirichlet $d$ brane
which is bounded by two NS branes in one direction. When the Dd brane
ends on the NS brane there is a local bending of the NS brane which
goes like $r^{d-4}$, $r$ denoting the radial coordinate in the
subspace of the NS world volume transverse to the Dd brane.  We can
measure the asymptotic bending by going to large $r$, along the world
volume of the NS brane. On the other hand, the transverse distance
between the two NS branes corresponds to the gauge coupling of the
system. The asymptotic bending of the NS branes controls the evolution
of the gauge coupling with scale.  Thus, when there is bending
which moves the two NS branes towards each other, the coupling grows
at high energies and we are dealing with a IR free theory.  When the
bending is outwards, the coupling becomes weak in the UV and we are
dealing with an asymptotically free theory.  Let us check what happens
for various dimensions. From the local bending behavior, we see that
for $d>4$ there is an asymptotic bending of the NS brane.  For $d<4$
there is no asymptotic bending. For $d=4$, the bending is logarithmic
and causes the NS brane to bend away from the D brane \cite{witMa}.

 From a field theory point of view, it is known that all gauge
theories in dimensions $d<4$ are asymptotically free, and that all field
theories with finite gauge couplings are infrared free for $d>4$; thus
we get perfect agreement with the brane picture.  For $d=4$, the
situation is more interesting, because a given NS brane may bend to
the left (right) at long distances if the number of D4 branes
intersecting it from the right is greater (less) than the number of D4
branes intersecting it from its left --- that is, if its linking
number \cite{hw} is positive (negative).  A given set of parallel D
branes between two NS branes will represent an asymptotically free
gauge group if the distance between the two NS branes grows with $r$,
an infrared free theory if it shrinks, and a finite theory if the two
NS branes are parallel at large $r$.  Thus, the relative distances
between NS branes far from the D4 branes reflects the beta functions
of the gauge groups lying between them, giving an intuitive correspondence
between bending and the beta function.  In particular, when all
of the NS branes are unbent, parallel to each other, and perpendicular
to the D4 branes --- which requires that the number of D4 branes
be the same in all the intervals between NS branes ---
then all the beta functions are  vanishing and the corresponding
theory is finite. Such models were discussed in \cite{witMa}.

Unfortunately, when the branes are rotated, as in \cite{EGK}, breaking the
supersymmetry down to \none\ in four dimensions, the bending of the NS
branes does not represent the beta function any more. Instead, the
bending of the branes corresponds to the R charges of the various
fields; see for example \cite{gkreview}.

However, in our construction, where we represent \none\
four-dimensional models using D5 branes on a two-dimensional interval,
the bending and beta functions are again related.  As described in
\cite{hz}, all the configurations drawn in this paper are given for
zero string coupling. When the string coupling is non-zero the branes
start to bend. The problem of bending is not fully solved, but we can
make some simple statements.

First, we can discuss simple configurations in which bending will not
be present.  In analogy to the case just discussed, the NS branes will
not bend at all if the number of D5 branes to their right and left are
everywhere equal.  A similar statement applies to the NS$'$ branes.
Clearly, for there to be no bending anywhere, the number of D5 branes
in every box should be the same.  As we will see below, this is indeed
the condition for a finite \none\ gauge theory!  Unlike the case of
eight supersymmetries, however, this is not a straightforward remark.
With eight supersymmetries, the finiteness of the theory is determined
by the one-loop beta functions only.  With four, one must consider the
interactions in the superpotential more carefully and ensure that the
conditions discussed in \SEC{fieldtheory} are satisfied.  It 
turns out that the brane configuration automatically adds the
correct superpotential and satisfies these conditions,
as is proven in Appendix B and discussed in the next section.

Second, it is easy to show that, far from the D5 branes, the NS and
NS$'$ branes tend to bend toward (away from) each other for infrared
(asymptotically) free theories.  For example, consider the case where
$n_{2,2}=N$ and $n_{1,1}=n_{1,2}=n_{2,1}= n_{2,3}=n_{3,3}=n_{3,2}= p$.
The $(2,2)$ box then corresponds to an \none\ $SU(N)$ gauge theory
with $3p$ flavors, with one-loop beta function coefficient $b_0 =
3(N-p)$.  Meanwhile, if $p>N$ ($p<N$), then at each edge of the $2,2$
box the number of D5 branes outside the box is greater (less than) the number
inside, so the NS and NS$'$ branes bordering the $2,2$ box will bend
inward (outward).  Thus the bending of the branes is sensitive to the
beta function.  However, in situations which are less symmetric, the
bending of the branes is not well understood.

To conclude, we propose a very simple criterion for a finite theory.
Given a brane configuration which has no bending, the corresponding
field theory which is read off from the brane configuration by using
the rules of \cite{hz} is a finite theory.  This criterion is a simple
generalization of the concept for theories with 8 supercharges, and
gives us a large class of finite $N=1$ theories with almost no
effort. For the cases where NS and NS$'$ branes do bend, we do not yet
fully understand the situation, but clearly the behavior of the branes
is related to the \none\ beta functions.

\subsection{$U(1)$ factors, Fayet-Iliopoulos terms, and flat directions}
\label{ssec:FIterms}

We now comment on Fayet-Iliopoulos (FI) parameters and the possible
``frozen'' $U(1)$ fields.  The gauge groups for zero string coupling
contain $U(1)$ factors.  It is argued in \cite{witMa} that these
$U(1)$ fields are frozen from a four dimensional point of view once
the string coupling is turned on.  We expect a similar situation here.
However, for the classical theory, with vanishing string coupling, the
$U(1)$ fields are present, and with them, FI parameters.

First we need an expression for the FI parameters. This is determined
as follows.  For $k=1$ or 0 and any $k'$, the system reduces to a
$N=2$ system and we can refer to previous results which give the FI
coupling for the $j$-th $U(1)$ $r_j=x_5^j-x_5^{j-1}$. Here $x_5^j$ is
the position of the NS$'$ brane in the 5 direction.  Equivalently, for
$k'=1$ or 0 and any $k$, we have again an $N=2$ system and the FI term
for the $i$-th $U(1)$ is $r_i=x_7^i-x_7^{i-1}$, with $x_7^i$ the
position of the $i$-th NS brane.  An expression which is consistent
with these two boundary conditions and assumes a linear relation gives
for the FI coupling of the $i,j$-th gauge group
$r_{i,j}=x_5^j-x_5^{j-1}+x_7^i-x_7^{i-1}$.

We can further check this proposal for the FI couplings by considering
the simple breaking of gauge symmetries generated by sending one NS
brane to far infinity in the 7 direction and reproducing the same
formulas for couplings.  The associated FI terms lead to the correct
breaking pattern in the classical field theory.  In the quantum
theory, the $U(1)$ fields will all be frozen, so instead the position
of the NS brane gets the interpretation of a baryonic expectation
value.  However, the pattern of symmetry breaking is the same.

\section{Marginal operators}
\label{sec:margops}

In this section we first explain the conditions for field theories
to have dimensionless couplings and associated marginal
operators, and then give a number of examples.

Although we will be interested in models with $k\times k'$ boxes on a
torus, it is easiest to begin our investigations by considering an
infinite grid of NS and NS$'$ branes, with $n_{i,j}\geq 2$ D5 branes
filling the square $(i,j)$. The field theory defined by this infinite
arrangement has at least one marginal coupling if, for every $2\times
2$ block of squares, the following condition holds:
\bel{crosscheck}
n_{i,j}+n_{i+1,j+1}=n_{i+1,j}+n_{i,j+1} \ .
\ee
The proof of the above claim is given in Appendix B.

Note that these (generically) chiral field theories are always anomaly
free.  For any square $(i,j)$, the number of fields in the fundamental
representation is $n_{i,j+1}+n_{i+1,j}+n_{i-1,j-1}$, while the number
in the antifundamental representation is
$n_{i,j-1}+n_{i-1,j}+n_{i+1,j+1}$.  The conditions \eref{crosscheck}
above ensure these are equal.  Furthermore, the condition ensures that
the difference between the number of D5 branes on the two sides of an
NS or NS$'$ brane remains a constant as one moves along the brane.
This ensures that the amount of bending which the NS or NS$'$ brane
undergoes remains constant as well.\footnote{The connection between
anomalies and bending was pointed out in \cite{gimgrem}.}

If we want a theory with no divergences, however, we must require as
an additional condition that all of the one-loop beta functions
vanish.  This demands that
\bel{finitecheck}
n_{i,j+1}+n_{i+1,j}+n_{i-1,j-1}=3n_{i,j}
\ee
for every $i,j$, whose only solution is $n_{i,j}=N$ for all $N$.  This
ensures that in finite models the NS and NS$'$ branes are unbent, as
noted earlier in \SSEC{bending}.

In addition, the finite models have still more dimensionless
couplings.  As shown in Appendix B, there will be one for each row
of boxes, one for each column of boxes, and one for each diagonal line
running from lower-left to upper-right (that is, along the $D$
fields) passing through the  boxes.

As described in \SEC{branes}, the models we want to study lie not on
an infinite plane but on a finite torus.  If we take a 
model on the $x_4-x_6$ plane $\IR^2$  whose grid of boxes and
choices of $\{n_{i,j}\}$ are left invariant by
a discrete lattice group $\Lambda$ of translations, we may consider
the theory on the torus formed by the quotient of the plane by
$\Lambda$.  The resulting theories all have a marginal coupling
associated with the overall K\"ahler parameter $\rho$ of the torus.
In addition, there may be additional marginal couplings in special
cases.  In the particular case where all boxes have $N$ D5 branes, we
find a {\it finite} theory, which in addition to the marginal coupling
associated with $\rho$ has marginal couplings for every
independent row, column, and lower-left/upper-right diagonal line
passing through the boxes.

 We will now give some examples of such models.

\subsection{\nfour\  $SU(N)$  gauge theory}

\begin{figure}
\centering
\epsfxsize=3in
\hspace*{0in}\vspace*{.2in}
\epsffile{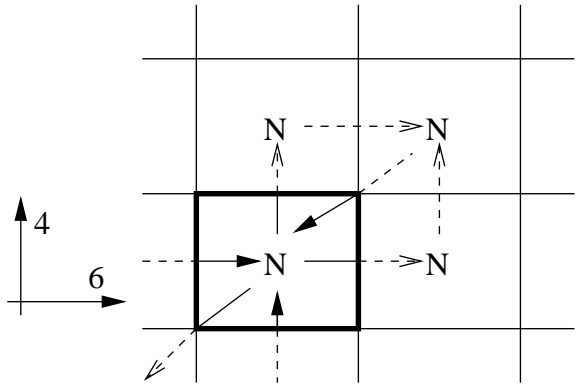}
\caption{The \nfour\ gauge theory; the torus consists of the one
highlighted box, of which the other boxes are images.  The three
arrows correspond to the three chiral multiplets in the adjoint
representation.}
\label{fig:onebox}
\end{figure}

First consider $k=k'=1$, as shown in figure \ref{fig:onebox}.  This
choice gives a single gauge group $SU(N)$ with \nfour\ supersymmetry.
The three fields $H, V$ and $D$ are charged under the same gauge group
because of the periodicity of the torus in the $x_4,x_6$ directions.
Being bi-fundamentals under the same group, they are adjoint fields
$\Phi_i$, $i=1,2,3$, giving the matter content of a \nfour\ theory.
(Again, classically these fields are adjoints of $U(N)$, and so
contain a gauge singlet for each of $H$, $V$ and $D$; however these
fields simply become the $x_7$ ($x_5$) position of the NS  (NS$'$)
brane in the quantum theory.)  Furthermore, the superpotential
\eref{HVD} in this case reads $W= H[V,D]$,
reproducing the appropriate \nfour\ interaction.  Note the importance
of the relative minus sign in \eref{HVD} between upper and lower triangle
contributions.

The gauge coupling,
equation \eref{gauge}, takes the form
\be
{1\over g^2} = {R_4R_6\over g_s l_s^2}.
\ee
and the theta angle, \Eref{theta}, is simply given as the
integral of the RR $B$ field on the torus
\be
\theta=\int_{46} B.
\ee
The complex gauge coupling constant $\tau$ is given by the combination of
these parameters, which corresponds to the K\"ahler parameter of the torus. In
the following we denote it by $\rho$ to avoid confusion with the
traditional name for the complex structure parameter of a torus.

Note that removing the NS and/or the NS$'$ brane from the torus does
not change the group; it merely corresponds to changing the
expectation value of the neutral scalars as discussed in
\SSEC{FIterms}.  Without the NS and NS$'$ brane, the D5 brane wrapped
on the torus is by T duality equivalent to a D3 brane, and the
K\"ahler parameter of the torus becomes the IIB string coupling
constant. This is consistent with the construction of \nfour\ gauge
theories from D3 branes, and provides a check on our construction.

\subsection{Elliptic \ntwo\ models}

\begin{figure}
\centering
\epsfxsize=5in
\hspace*{0in}\vspace*{.2in}
\epsffile{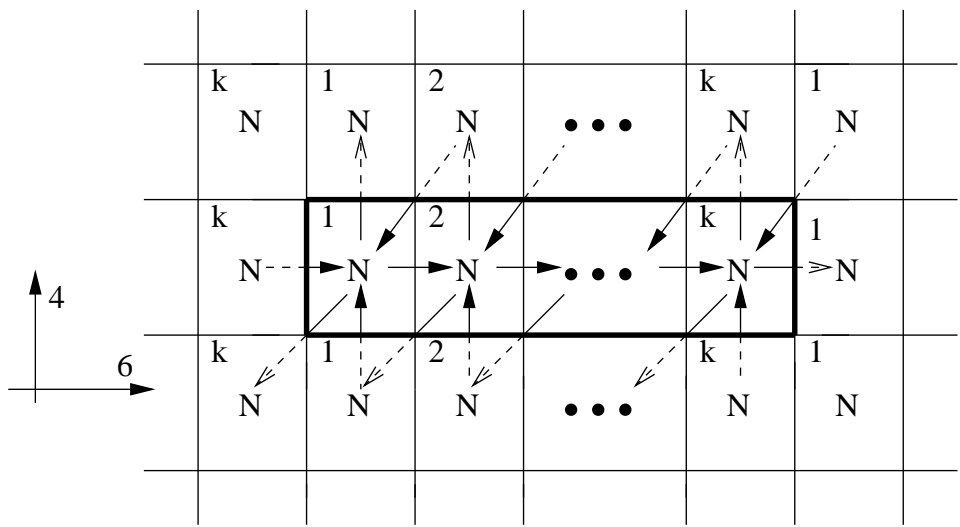}
\caption{Construction of the \ntwo\ elliptic models. The theory has
gauge group $SU(N)_1\times SU(N)_2\times \cdots \times SU(N)_k$; each
box corresponds to one of these $SU(N)$ factors, as indicated by the
indices in the upper left corners.}
\label{fig:kbox}
\end{figure}

Consider a model on a torus identified under vertical shifts by one
box and under horizontal shifts by $k$ boxes.  Each of the $k$ boxes
contains $N$ D5 branes, as shown in figure \ref{fig:kbox}.  (The
construction has $k$ NS branes and one NS$'$ brane; however, as in the
previous case, the one NS$'$ brane may be removed without changing the
theory.)  This is an \ntwo\ \susic\ $SU(N)\times
SU(N)\times\cdots\times SU(N)$ gauge theory, with hypermultiplets in
bifundamental representations.  The \none\ chiral multiplet in the
\ntwo\ $SU(N)_i$ vector multiplet is the field $V_{i,1}$, while the
hypermultiplets under $SU(N)_i\times SU(N)_{i+1}$ are the fields
$(H_{i,1},D_{i+1,1})$. The \none\ superpotential for this model also
reproduces the correct \ntwo\ interactions.  The model has $k$
dimensionless couplings, corresponding to the overall size of the
torus and to $k-1$ motions in $x_6$ of the NS branes (these couplings
are complexified by the parameters from the Wilson lines of the NS
brane world-volume gauge fields, see Appendix A.)  These models are
the same as the elliptic models of \cite{witMa}, to which they are
related by T duality in the $x_4$ direction.

These models have flat directions (classically they are
Fayet-Iliopolous parameters of the $U(1)$ groups which are absent
quantum mechanically) which break the $k$-box model to the $k-1$ box
model; these motions correspond to moving one of the NS branes off of
the D5 branes.  If all but one or zero of the NS branes is removed,
the resulting 1-box model is the \nfour\ gauge theory considered
above.

\subsection{Elliptic \none\ models}

A number of elliptic finite \none\ \susic\ models may be constructed in
this way, by considering a torus of $k\times k'$ boxes with each
box filled with $N$ D5 branes. In this section we consider a
few examples.

Consider first the $k=k'=2$ model of figure \ref{fig:twobytwo}.  This
theory is vectorlike: there is a bifundamental and its complex
conjugate connecting every pair of groups, for a total of twelve
fields. It is easy to check this model is finite, and that it has a
total of four dimensionless couplings: one for the overall torus, one
for the separation of the NS branes, one for the separation of the
NS$'$ branes, and one for the diagonal.

\begin{figure}
\centering
\epsfxsize=4in
\hspace*{0in}\vspace*{.2in}
\epsffile{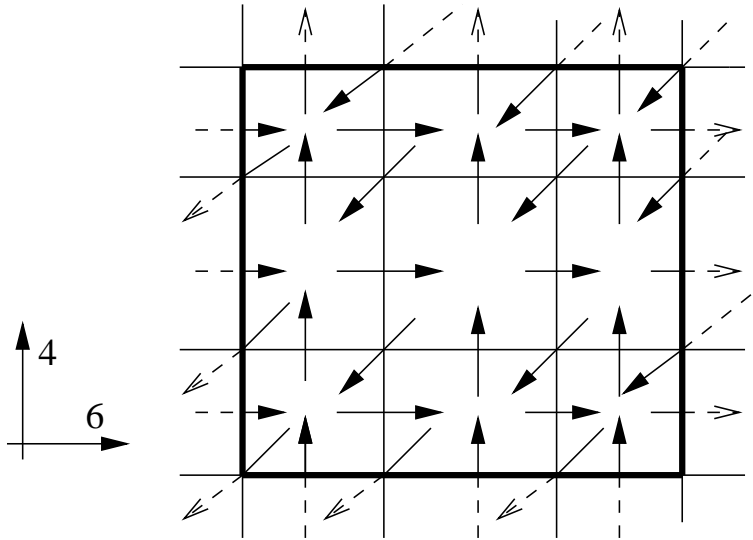}
\caption{Construction of the $3\times 3$ $[SU(N)]^9$ chiral model.}
\label{fig:threebythree}
\end{figure}

Next consider a $k=k'=3$ model, as in figure \ref{fig:threebythree},
where the grid of boxes is identified under shifts by three
boxes vertically and by three boxes horizontally.
This model has eighteen fields in chiral representations; there is a
bifundamental {\it or} a biantifundamental connecting each pair of
groups except those whose boxes lie in an upper-left/lower-right
orientation.  There are seven dimensionless couplings: two from rows,
two from columns, two from diagonals, and one from $\rho$.

Each of these models has flat directions (or Fayet-Iliopolous
parameters) along which it flows to a model with fewer boxes by
removal of NS or NS$'$ branes.  For example, one may have the
transitions $3\times 3 \rarr 3\times 2 \rarr 2\times 2\rarr 2\times
1\rarr 1\times 1$ in which the chiral \none\ model flows to a
vectorlike \none\ model, which then flows to an \ntwo\ model and from
there to an \nfour\ gauge theory.

\begin{figure}
\centering
\epsfxsize=4.5in
\hspace*{0in}\vspace*{.2in}
\epsffile{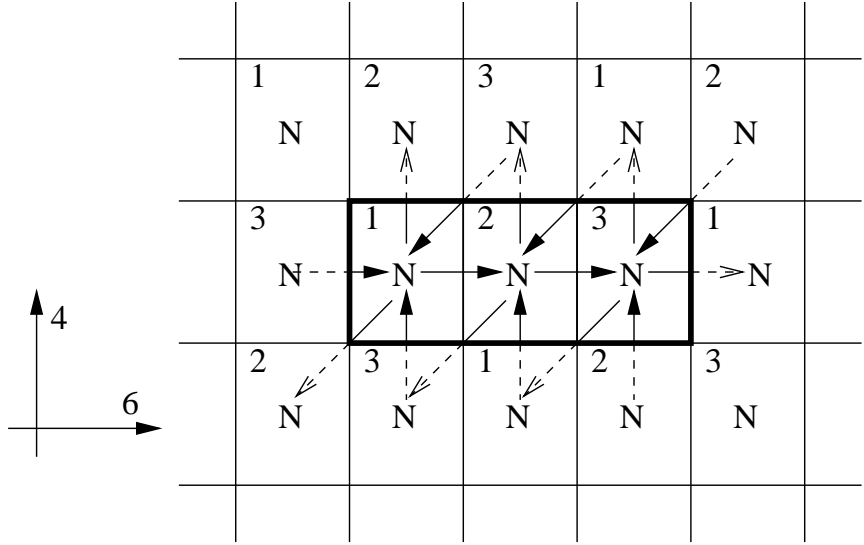}
\caption{Construction of the $1\times 3$ twisted model.  The theory is
$SU(N)_1\times SU(N)_2\times SU(N)_3$; each box
corresponds to one of these $SU(N)$ factors,
as indicated by the indices in the upper left corners.}
\label{fig:Zthree}
\end{figure}

Next consider a $1\times 3$ model, as in figure \ref{fig:Zthree}, where
the torus is obtained by quotienting an infinite grid by the translation
group generated by a horizontal translation (of three boxes) and a
{\it diagonal} translation (of one box up and one box to the left).
Equivalently, this is the previous $3\times3$
model modded out by a freely acting $Z_3$ transformation.
The spectrum contains three copies of chiral fields transforming as
follows under $SU(N)_1 \times SU(N)_2 \times SU(N)_3$
\be
Q_{1a}=(1,\fund,\antifund) \;\; ; \;\; Q_{2a}=(\antifund,1,\fund)
\;\; ; \;\; Q_{3a}=(\fund,\antifund,1)
\ee
with $a=H,V,D$ for horizontal, vertical and diagonal fields. The
superpotential can be recast as the simple expression
\be
W=\epsilon^{ijk} Q_{iH} Q_{jV} Q_{kD}
\ee
Here there is only one dimensionless coupling; there is only one NS
brane and one NS$'$ brane, and so only the overall size of the torus
plays a role. The three gauge couplings are fixed to be equal.  (Note
that, while this is strictly true of the brane construction for
geometrical reasons, the field theory is more general, and only need
have equal couplings when the condition of finiteness is imposed.)
This model has recently been discussed in \cite{Ibanez,ks}, where it
appeared in the worldvolume of D branes at $\IC^3/Z_3$
singularities. It would be interesting to clarify the relation between
both approaches.  Clearly, many generalizations of this model are
possible.

\subsection{Cylindrical \ntwo\ and \none\ models}

If we allow the torus to degenerate to a cylinder,
by allowing the area of the torus to become infinite
while keeping $R_4$ or $R_6$ finite, we find a new
class of models as a limit of our elliptic models.

\begin{figure}
\centering
\epsfxsize=4in
\hspace*{0in}\vspace*{.2in}
\epsffile{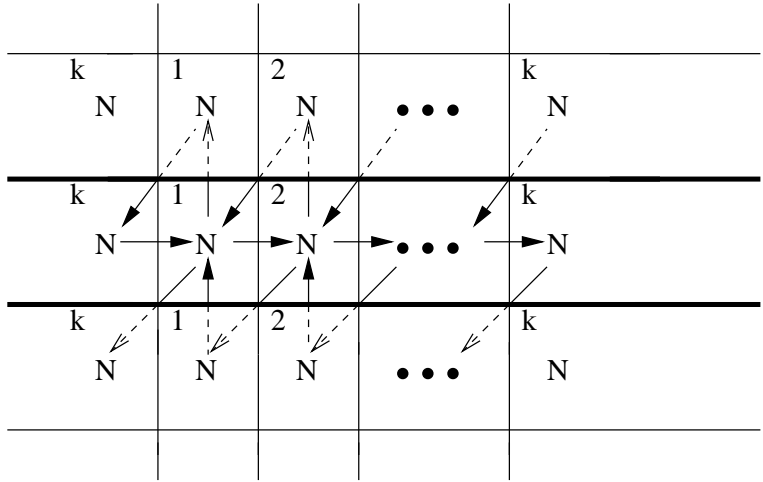}
\caption{Construction of the \ntwo\ cylindrical model with gauge group
$[SU(N)]^{k-1}$. The cylinder is shown as a lattice modulo a unit
cell, whose boundary is indicated with a heavy outline.}
\label{fig:ntwocyl}
\end{figure}

For illustration we begin with the \ntwo\ case, which is discussed in
\cite{witMa}.  If $k=2,k'=1$ in figure \ref{fig:kbox}, we have an
\ntwo\ gauge theory with $SU(N)_1\times SU(N)_2$ as gauge group.  We
may imagine letting the $x^6$ length of the second box grow to
infinity while that of the first box remains finite; thus, with the NS
branes held at a fixed distance $L_6$, and with $R_4$ held constant,
$R_6$ goes to infinity.  We are left with a cylinder of radius $R_4$
with one finite box and two infinite boxes (both corresponding to
$SU(N)_2$ on either side).  $SU(N)_2$ becomes free in this limit and
becomes a global symmetry of $SU(N)_1$; we are left with $SU(N)_1$
with $2N$ hypermultiplets, a well-known finite theory with duality
\cite{nsewtwo,sundual,sundualb}.  Similarly, using larger values of
$k$, we may construct finite \ntwo\ theories with $[SU(N)]^{k-1}$
groups, which have a row of $k-1$ boxes of finite size and boxes on
either end of infinite extent in the $x_6$ direction.  This is shown
in figure \ref{fig:ntwocyl}.

\begin{figure}
\centering
\epsfxsize=3in
\hspace*{0in}\vspace*{.2in}
\epsffile{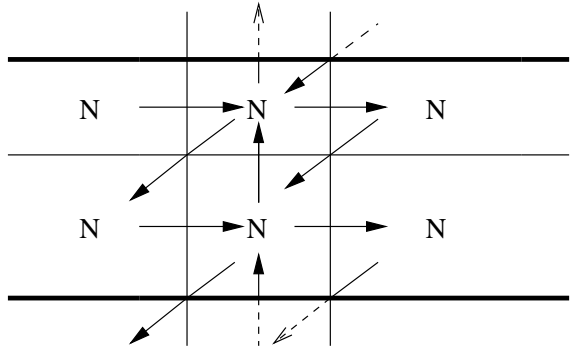}
\caption{Construction of the \none\ cylindrical model
with gauge group $SU(N)\times SU(N)$.}
\label{fig:nonecyl}
\end{figure}

More generally, in any elliptic model whose symmetries do not forbid
it, we may take the limit where the torus becomes a cylinder while a
set of rows or of columns of boxes retains finite area.  For example,
we may take the model of figure \ref{fig:twobytwo} in the limit that
the $x_6$ extent of the boxes $(2,1)$ and $(2,2)$ goes to infinity
(figure \ref{fig:nonecyl}.  The boxes in the left column remain
interacting, while the right column and its image to the left of the
left column become non-interacting flavor groups of the central
column.  The resulting \none\ $[SU(N)]^2$ model is finite.  There is a
bifundamental and biantifundamental of the $SU(N)\times SU(N)$ group,
and each group factor also has $2N$ flavors from the adjacent flavor
groups.  Note that all the fields which are neutral under the gauge
group decouple from the theory.  This model has a flat direction,
corresponding to the removal of an NS$'$ brane from the grid, along
which it flows to the \ntwo\ $SU(N)$ with $2N$ hypermultiplets.  In a
similar way we may obtain finite cylindrical $[SU(N)]^{(k-1)k'}$
theories from elliptic $[SU(N)]^{kk'}$ models.

 These cylindrical models obviously have fewer marginal couplings than
those of the elliptic models of which they are a limit.  In particular
the K\"ahler parameter $\rho$ of the torus does not exist for the
cylinder.

\subsection{Non-finite models with marginal operators}

These models are interesting (and to our knowledge have not appeared
before) because they contain (1) asymptotically free subgroups, (2)
infrared free subgroups, and (3) a completely renormalizable
superpotential.  The brane picture is less well understood in this
case; the branes bend in complicated ways which as yet we do not know
how to characterize properly.  However, the geometrical simplicity of
the brane construction is still useful for building these field
theories.

\begin{figure}
\centering
\epsfxsize=2.5in
\hspace*{0in}\vspace*{.2in}
\epsffile{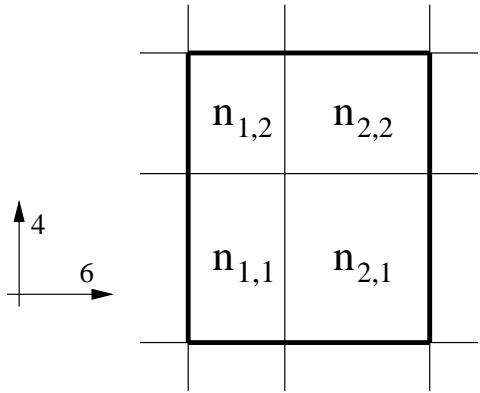}
\caption{Construction of a non-finite model with a marginal operator;
here $n_{1,1}+n_{2,2}=n_{1,2}+ n_{2,1}$.}
\label{fig:mtwobytwo}
\end{figure}

The simplest such model is the torus with $2\times 2$ boxes (with
identifications under shifts of two boxes vertically and two boxes
horizontally) in which the number of D5 branes is $n_{1,1}=n_{2,2}=N$,
$n_{1,2}= N+k$, $n_{2,1}= N-k$.  The matter content is as in
\Eref{fourboxmatter}.  The $SU(N-k)$ group factor has a positive
one-loop beta function, the $SU(N+k)$ group factor has a negative
one-loop beta function, and so the theory is not finite and indeed is
non-renormalizable.  However, in the limit that the gauge couplings
for all groups except $SU(N+k)$ are zero, the $SU(N+k)$ theory has a
conformal fixed point at low energy \cite{NAD}.  Its matter fields
have non-trivial anomalous dimensions.  By turning on the gauge
couplings of the other factors, we can deform this fixed point
continously \cite{emop}.  This fact cannot be seen as yet in the brane
picture; our understanding of the bending of the branes is still
incomplete.  Nonetheless, the field theory story is clear.  In fact,
it could be studied in detail using the fact that it has a large $N$
limit; for $N$ large and $k$ small, this marginal line begins in the
perturbative regime.  The theory has two marginal couplings, one from
the overall scale of the torus and one from the lower-left/upper-right
diagonal line (see Appendix B.)

More generally we can consider a $2\times 2$ model with arbitrary
$n_{i,j}$ satisfying $n_{1,1}+n_{2,2}=n_{1,2}+ n_{2,1}$.  This model is
vectorlike and can have a one-dimensional manifold of fixed points.
The dynamics may not always permit this, however.  For example, if
$n_{1,1}=n_{2,1}=9$ and $n_{2,1}=n_{2,2}=2$, then the $SU(9)$ groups have
13 flavors and the $SU(2)$ groups have 20 flavors.  If we set all but
the $n_{1,1}$ coupling to zero, the $SU(9)$ theory with 12 flavors is
in the free magnetic phase, and therefore does not reach a fixed
point.  Indeed a fixed point of this type would be inconsistent
with unitarity \cite{NAD}.  Thus one must use this construction
with some care; the semiclassical geometry alone is not sufficient.

A similar issue prevents the existence of non-finite \ntwo\ elliptic
models.  For example, consider the torus with two boxes containing
$N_1$ and $N_2$ D5 branes.  If $N_1>N_2$, then the first factor is
asympotically free while the second is infrared free.  If we turn off
the coupling of the second, the first has no fixed point in the
infrared corresponding to an unbroken $SU(N_1)$ gauge theory.  (There
may be Argyres-Douglas fixed points \cite{ArDg} on the moduli space,
but these are not governed by naive application of the formulas
\eref{betah}-\eref{betaSV}; see \cite{APSW}.)  Thus there are no
solutions to the requirement that all beta functions vanish.

\begin{figure}
\centering
\epsfxsize=5in
\hspace*{0in}\vspace*{.2in}
\epsffile{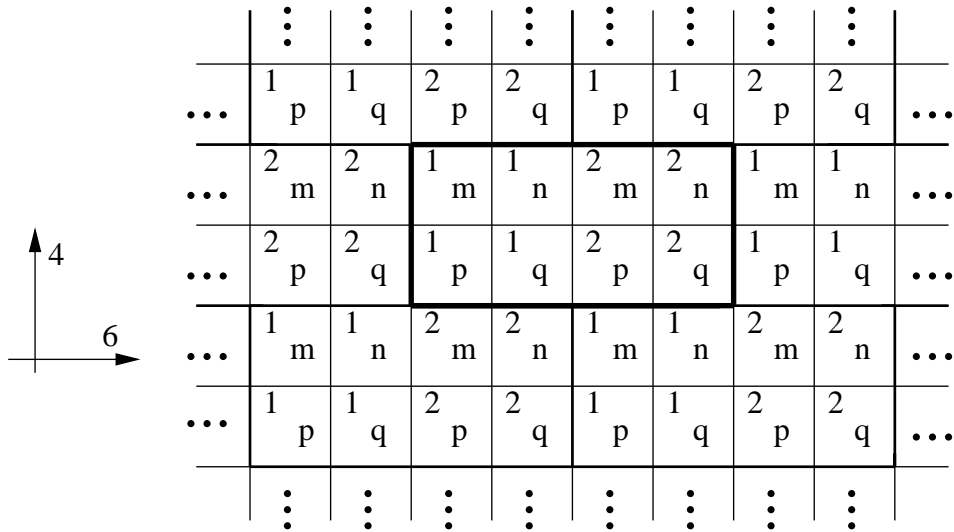}
\caption{A twisted non-finite model with at least one marginal
operator.  The theory is $SU(m)_1\times SU(n)_1\times SU(p)_1\times
SU(q)_1 \times SU(m)_2\times SU(n)_2\times SU(p)_2\times SU(q)_2$;
each box corresponds to one of these factors, as indicated by
the indices in the upper left corners.}
\label{fig:twobyfour}
\end{figure}

We may also construct chiral models of this type by taking $k$ or $k'$
greater than two. For a twisted example, consider the case $k=4$,
$k'=2$ where the torus is identified as a grid modulo shifts of four
boxes to the right and of two boxes diagonally to the upper right, as
in figure \ref{fig:twobyfour}.  As long as the two $2\times 2$ blocks
have the same gauge groups (with \Eref{crosscheck} satisfied), the
model will be well defined and have a marginal operator.  Note also
that it has a large $N$ limit; if $m,n,p,q$ are of order $N$ while the
differences between them are order 1, then the manifold of fixed
points of this theory will be visible in perturbation theory.

Many of these theories have flat directions (classically,
Fayet-Iliopolous terms) along which they flow to theories with fewer
boxes.  To remove a single NS or NS$'$ brane we must have the same
number of D5 branes on either side of each portion of the NS or NS$'$
brane.  In these models, where the numbers of D5 branes may vary from
box to box, this condition is generally not satisfied.  However, in
special cases various branes may be removed; for example, if $n_{i,j}=
n_{i+1,j}$, then the brane between the $i$ and $i+1$ columns is not
bent and may be lifted off the D5 branes.  But there are other flat
directions, involving the removal of pairs of NS or NS$'$ branes,
which certainly arise.  For example, if $n_{i,j} = n_{i+1,j}+p =
n_{i-1,j}+p$ for all $j$, then one may remove the $i^{th}$ and
$(i-1)^{th}$ NS branes along with $p$ D5 branes (which wrap the torus
in the $x_4$ direction but are finite in $x_6$), leaving behind boxes
with $n_{i-1,j}$ D5 branes.  This corresponds to breaking the theory
to a $k-1\times k'$ box \none\ model times a pure \ntwo\ $SU(p)$ gauge
theory.

Finally, let us note that in special cases one may construct
interesting non-finite cylindrical \none\ models in analogy to the
cylindrical models of the previous section.    

The construction of similar \none\ theories which are not finite but
have at least one marginal coupling is straightforward.  We have not
worked out the general counting rules for the number of such
couplings, but in any given case these can be worked out using the
field theory rules \eref{betah}-\eref{betaSV}.  Much remains
to be understood in the brane picture, including the
full story of the bending of the branes and the counting
of marginal couplings.

\subsection{Still more non-finite models}

From certain models --- for example, those with $k=2$ --- additional
non-renormalizable models with quartic superpotentials and marginal
couplings, such as in \Eref{sufourB}, can be derived.  For $k=2$, the
fields $H_{1,j}$ and $H_{2,j}$ are complex conjugates; consequently, a
mass term can be written for them, although we do not know how to
represent this using branes.  In field theory, integrating out these
massive fields leads to a gauge theory with the quartic superpotential
\be
W={1\over M}\sum_{j,J}\left[ h_{1,j}^+ V_{2,j}D_{2,j+1}
- h_{1,j-1}^- V_{1,j-1}D_{2,j}\right]
\left[h_{2,J}^+ V_{1,J}D_{1,J+1}
-h_{i,J}^-V_{2,J-1}D_{1,J}\right].
\ee
which, by construction \cite{emop}, may have dimensionless couplings in its
infrared conformal field theory.  It is natural to expect \SL\ and its
generalizations to act on these couplings, as in certain previously
known cases \cite{emop,ails}.  Note that this construction applies
independent of whether the original model is finite.

\section{Duality}
\label{sec:duality}

We consider some examples of the previous section, discussing
their duality properties in turn.

\subsection{The \nfour\ theory}

The simplest case is the \nfour\ $SU(N)$ theory, which is realized in
our framework as $N$ D5 branes wrapping the $T^2$ in the 46
directions. As mentioned above, we can remove the NS and NS$'$
branes in this case without changing the theory.

Recall that the gauge coupling constant is given by $1/g_{YM}^2= (R_4
R_6)/(g_s l_s^2)$, and the theta angle is given by the integral of the
RR 2-form over the $T^2$. Both parameters are arranged in the K\"ahler
parameter $\rho$ of the torus. In this toroidal compactification of
the type IIB string theory there is a natural \SL\ action on $\rho$
\cite{grp} that leaves the physics invariant, and which corresponds to
\OM\ duality in the \nfour\ gauge theory.  It is interesting to
compare this construction to other realizations of the \nfour\ theory
via branes. After T duality along 4 and 6, the D5 branes become D3
branes sitting at a point in the dual $T^2$. The geometric factors in
$1/g_{YM}^2$ are absorbed in the redefinition of the string coupling;
one then has $1/g_{YM}^2=1/g_s'$.  The integral in 46 of the RR 2-form
transforms into the RR scalar.  The complex gauge coupling in the
gauge theory on the world-volume of the D3 branes is thus given by the
type IIB complex coupling constant theory.  Thus the T duality maps
the K\"ahler parameter $\rho$ of the initial torus to the string
coupling constant in the T dual theory. The \SL\ duality of ten
dimensional type IIB string theory \cite{ht,witdim} accounts for the
\OM\ duality on the D3 brane world volume \cite{tseytlin,gregut}.

Another construction of this theory, which is more convenient for our
purposes, is obtained by going to an M-theory description. We can
perform a T duality along one circle, say $x^4$; let $x^{4'}$ be
the coordinate of the dualized circle. The gauge
theory is now realized on the world-volume of $N$ type IIA D4 branes wrapped
on the $x^6$ circle, as in \cite{witMa}.  We then treat type IIA on the
$T^2$ parametrized by $x^{4'},x^6$ as M theory on a $T^3$; the D4
branes become M5 branes wrapped on a two torus parametrized by
$x^6,x^{10}$. Denoting by $R_{4'}, R_6$ and $R_{10}$ the lengths of
circles in M-theory, the standard relations following from
M-theory/type IIB duality \cite{aspinwall,schwarz} are (up to
numerical factors) \be R_4 = \frac{l_{11}^3}{R_{4'} R_{10}} \;\;\; ;
\;\;\; g_s = \frac{R_{10}}{R_{4'}} \;\;\; ; \;\;\;
l_s^2=\frac{l_{11}^3}{R_{10}}\; .  \ee The gauge coupling in the type
IIB construction is now given by $1/g_{YM}^2=R_6/R_{10}$. The theta
angle in the type IIB construction $B_{46}$ becomes, under the T
duality, the component along 6 of the IIA RR vector field, $A_6$,
which corresponds to the $6, 10$ component of the M-theory
metric. These two parameters combine to define the complex structure
of the M-theory two-torus along $6, 10$. Thus, the transformation to an
M-theory description maps the K\"ahler parameter of the IIB torus to
the complex structure of an M-theory torus. As in \cite{witMa}, the
geometric \SL\ action on this last accounts for \OM\ duality in the
gauge theory.

\subsection{\ntwo\ models}

Let us now consider the \ntwo\ models which are realized by
introducing $k$ NS branes (or $k'$ NS$'$ branes) in the type IIB
construction, as in figure \ref{fig:kbox}. The torus along $46$ is
divided in $k$ boxes, each giving a $SU(N)$ gauge factor.  As already
stated, the individual gauge couplings and theta parameters are
encoded in the positions of the NS branes and the Wilson lines of
their world-volume $U(1)$ gauge fields along $x^4$. As mentioned earlier,
these models are the same as the elliptic models of \cite{witMa},
where their duality properties were discussed.  We review the
construction of the duality group that acts on the parameter space.

The space of parameters is easily visualized by going to the M-theory
description as we did just a moment ago. T dualizing along 4 and
lifting the IIA configuration to M-theory, the IIB D5 branes become M5
branes wrapping a $T^2$ parametrized by $x^6, x^{10}$ (denoted as $E$
in what follows), and the NS branes become M5 branes point-like in
these coordinates and spanning $4'$ and 5. The K\"ahler parameter of
the type IIB torus becomes the complex structure parameter of $E$, and
the positions and Wilson lines of the NS branes become the
positions of the corresponding $k$ M5 branes in 6,10. The space of
parameters is the moduli space of genus one smooth Riemann surfaces
$E$ with $k$ unordered marked points \cite{witMa}. It is denoted by
$\CM_{1,k}$. A point in $\CM_{1,k}$ defines a gauge theory with
overall coupling constant given by the complex structure on the
Riemann surface (equivalently, a parameter $\tau$ in the fundamental
domain of \SL), and individual gauge couplings defined by the
positions of the $k$ points on the Riemann surface. The fact that the
points are unordered reflects the fact that the NS branes are
indistinguishable.

\begin{figure}
\centering
\epsfxsize=5in
\hspace*{0in}\vspace*{.2in}
\epsffile{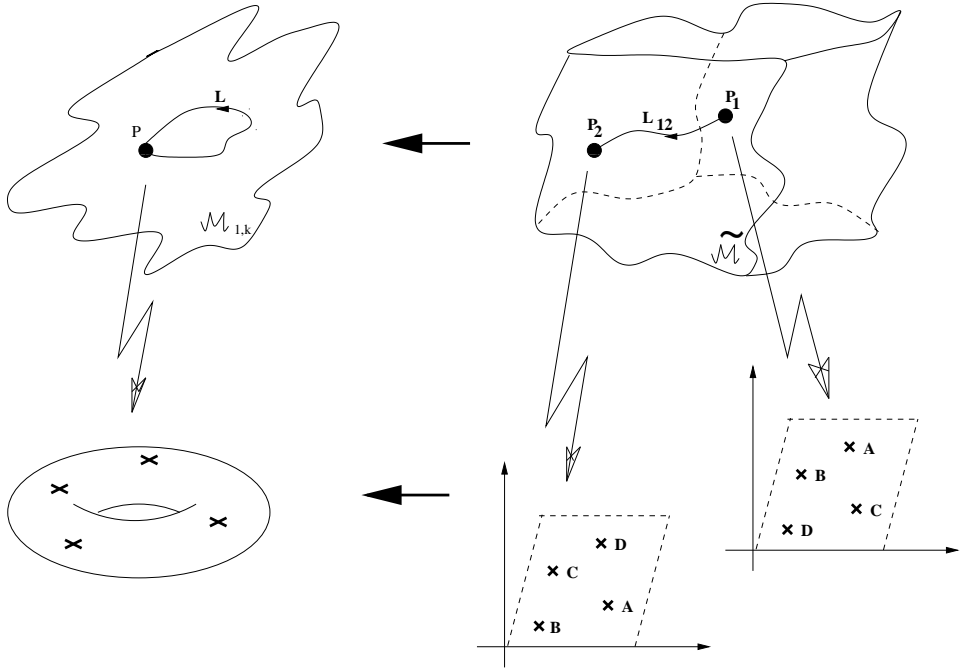}
\caption{An example of a duality operation: Each point in the
universal cover ${\tilde \CM}$ represents the choice of coordinates
for the M5 branes labelled A, B, C and D. The points $P_1$ and $P_2$
differ only in that the branes are permuted.
As the M5 branes are {\em physically} indistinguishable, the
two points $P_1$ and $P_2$ are a single point $P$ in the
space of {\em physical} parameters $\CM_{1,k}$.  A curve $L_{12}$
connecting $P_1$ and $P_2$ in ${\tilde \CM}$ is thus mapped to a
homotopically non-trivial closed loop $L$ in $\CM_{1,k}$.}
\label{fig:duality}
\end{figure}

Following \cite{witMa}, the duality group is associated to closed
loops in this space (changes in the parameters of the theory that
leave the physics invariant) with the equivalence relation defined by
smooth deformation (homotopy). So the duality group is
$\pi_1(\CM_{1,k})$ \cite{witMa}. The simplest way of analyzing what
type of duality transformations we are describing is to consider the
universal cover ${\tilde {\CM}}$ of $\CM_{1,k}$, and determine the
actions we have to quotient ${\tilde{\CM}}$ by to obtain
$\CM_{1,k}$. These actions constitute the duality group:
$\CM_{1,k}={\tilde \CM}/\pi_1(\CM_{1,k})$.

The space ${\tilde{\CM}}$ can be thought of as $\IC^+ \times
(\IR^2)^{k}$, where $\IC^+$ denotes the classical space of the
overall coupling constant, namely the upper half complex plane with no
\SL\ identifications. The $i^{th}$ factor $\IR^2$ corresponds to the
positions of the $i^{th}$ M5 brane on a complex plane. A
non-minimal set of generators of $\pi_1(\CM_{1,k})$ is given by
\begin{itemize}
\item \SL\ transformations on $\IC^+$.
\item Shifts in the $i^{th}$ $\IR^2$ by whole periods of the torus. In the
type IIB construction this corresponds to shifting the $x^6$ positions of
the $i^{th}$ NS brane by a multiple of $R_6$, or, similarly, shifting
its Wilson line along 4 by whole periods.
\item Permutations of the $\IR^2$ factors. This amounts to interchanging
the NS branes in the type IIB picture (as in figure \ref{fig:duality}.)
\end{itemize}

Thus the group contains in general not only the overall \SL\ but also
additional elements. When the NS branes are removed, leaving the
theory \nfour\ invariant, the overall \SL\ duality group of the
$k$-box model becomes the \SL\ duality group of the \nfour\ theory.

At this point, it is useful to analyze another M-theory description of
the same type IIB \ntwo\ brane configurations. This is obtained by T
duality along 6, and transforms the D5-branes into D4-branes wrapping
the $x^4$ circle, and the NS branes into Kaluza-Klein (KK) monopoles,
described by a multicentered Taub-NUT space.  Notice that the $k$
centers in the Taub-NUT space are coincident, since their $x^7, x^8$
and $x^9$ coordinates all vanish. The $x^6$ positions of the IIB NS
branes are encoded in the non-vanishing integrals of the IIA NS-NS
2-form on the $k$ 2-cycles $\Sigma_i$ of the KK monopole. Going to
M-theory by growing a compact dimension ${\overline{10}}$, the
D4-branes become M5-branes wrapping a two-torus along $4,{\overline
{10}}$ (denoted ${\overline E}$), of complex structure parameter
$\rho$ (the K\"ahler parameter of the IIB torus). The multi Taub-NUT
metric has translational invariance in ${\overline{10}}$, and in this
sense the KK monopoles wrap ${\overline E}$ as well. The parameters of
the IIB NS branes are encoded in the integrals of the M-theory 3-form
$C_3$ over the $k$ $\IP_1$'s in the Taub-NUT space times each of the
two independent circles in ${\overline E}$,
\be
\int_{\Sigma_i \times S^1_{(4)}} C_3 \;\;\; ; \;\;\;
\int_{\Sigma_i \times S^1_{({\overline{10}})}} C_3.
\ee
These may be called, in a broad sense, Wilson lines along
$4,{\overline {10}}$ of the gauge fields $\int_{\Sigma_i} C_3$. The
space of parameters is the moduli space of smooth Riemann surfaces
${\overline E}$ with a choice of $k$ indistinguishable pairs of these
Wilson lines. By comparison with the description of the same theory in
the previous paragraph, we learn that a choice of a pair of Wilson
lines on ${\overline E}$ corresponds to a choice of a point in
$E$. Equivalently, a pair of Wilson lines on ${\overline E}$ lives
(takes values) on a two-torus $E$ with the same complex structure. The
parameter space is then composed of choices of complex structure and
$k$ indistinguishable points in $E$, again $\CM_{1,k}$, and the
duality group is its fundamental group.  Notice that the compact
dimension $6'$ does not play any role in the determination of the
duality group.

\subsection{\none\ models}

Now we can turn to the analysis of the \none\ theories, obtained by
introducing $k$ NS branes and $k'$ NS$'$ branes. There is a difficulty
in the determination of the duality group in this case. In the \nfour\
and \ntwo\ theories, the brane construction displayed very explicitly
the marginal operators which were shown to exist using field theory
techniques. Thus the space of parameters had a simple geometrical
representation, which we uncovered by going to an M theory
description.  In the generic \none\ case, however, the field theory
analysis shows the existence of additional marginal operators (those
associated with diagonal lines of boxes) for which we have not found
an interpretation in terms of parameters in the brane
configuration. These parameters may be associated to the fields living
at the intersection of the NS and NS$'$ branes (as discussed in
Appendix A), even though the precise connection is not clear enough to
allow for, {\it e.g.}, matching the counting of these parameters in
the field theory and the brane construction. Another possibility is
that these parameters may not be realized in the brane configuration.

We leave the precise characterization of the complete parameter space
and duality group of these gauge field theories for future work, and,
in the following, restrict ourselves to discussing the subspace of
parameters which is manifest in the brane construction, and to
determining the subset of duality transformations acting on it.

We start with the IIB brane configuration with both NS and NS$'$
branes. Performing a T duality along 4, the D5 branes become M5 branes
wrapping the two torus $E$ in $6, 10$, the NS branes become M5 branes
pointlike in $E$, and the NS$'$ branes transform into KK monopoles
wrapping $E$.  The compact dimension $4'$ will not play any role.  The
space of parameters corresponds to choices of a complex structure for
$E$, of $k$ unordered marked points on $E$ (positions of NS branes)
and $k'$ unordered marked points on ${\overline E}$ (choice of Wilson
lines around $E$.) \footnote{In the \ntwo\ section we showed that
Wilson lines around ${\overline E}$ correspond to points in $E$; an
analogous argument shows that Wilson lines around ${E}$ correspond to
points in ${\overline E}$.} Recalling that $E$ and ${\overline E}$
have the same complex structure, and that their K\"ahler classes are
irrelevant, there is no obstruction to identifying them for the
purposes of computing the duality group. The space of parameters is
then the moduli space of a smooth Riemann surface with $k$ unordered
marked points of a certain kind, and $k'$ unordered marked points of
another. We denote this space by $\CM_{1,k,k'}$; the associated
duality group is $\pi_1(\CM_{1,k,k'})$.  The full duality group of the
theory will contain this as a subgroup.

From our experience with  \ntwo\ duality we see that the transformations
described include the overall \SL, shifts of $x^4$ positions and Wilson
lines of NS$'$ branes by whole periods, shifts of $x^6$ positions and
Wilson lines of NS branes by whole periods, and permutations of NS
fivebranes of the same kind.

Notice the symmetry between the two kinds of points, corresponding to
the equivalence of the M-theory descriptions obtained by T dualizing
along 4 or 6 in the initial IIB configuration. Notice also that
whenever any kind of NS branes is absent one recovers the duality
group for the \ntwo\ theory.

The description of the space of parameters presented above requires
further discussion when the fundamental translations defining the type
IIB torus are not simple horizontal and vertical shifts, as in the
model of figure \ref{fig:Zthree}. As argued above, there exist
relations in this case among the couplings of the different gauge
factors, and the space of parameters is smaller than what the number
of group factors would suggest.

\begin{figure}
\centering
\epsfxsize=3in
\hspace*{0in}\vspace*{.2in}
\epsffile{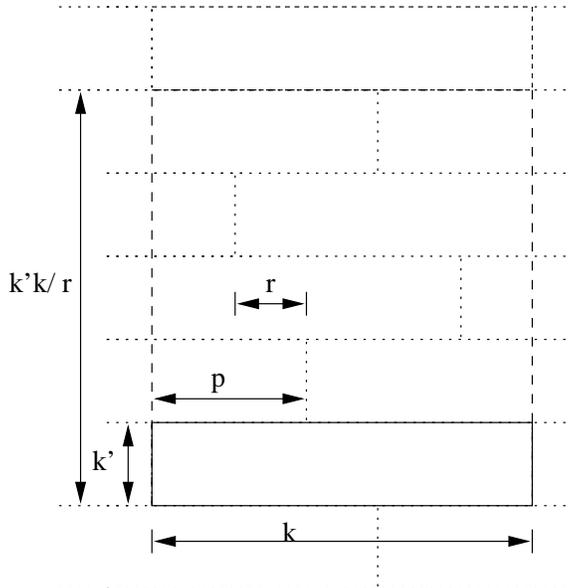}
\caption{Construction of the $k \times k k'/r$ box model from the $k
\times k'$ box model with non-trivial identifications. The unit cell of
the $k \times k'$ box model is the solid rectangle.
By adjoining a sufficient number of shifted rectangles (in dotted
lines) one can define a $k\times k k'/r$ cell (in dashed lines) whose
sides are identified in the trivial way. The $k\times k'$ box model is the
quotient of the $k\times k k'/r$ box model by a finite group $\ZZ_{k/r}$
of translations in the torus.}
\label{fig:twistdual}
\end{figure}

Without loss of generality we can consider a $k \times k'$ box model in
which the identification of the vertical sides of the unit cell is trivial
(defined by a horizontal translation), but the identification of
horizontal sides is accompanied by a horizontal shift of $p$ boxes.
Let $r$ denote the greatest common divisor of $p$ and $k$. The
theory can be considered as originating from a $k \times k k'/r$ box model
with trivial identifications, modded out by a freely acting $Z_{k/r}$
translation in the torus. This is schematically depicted in Figure
\ref{fig:twistdual}.

Let us consider the space of parameters of a $k \times k k'/r$ box
model, without any $Z_{k/r}$ identification. In the M-theory
description we have the torus $E$ with $k$ points on it. The
coordinates of these points correspond, when expressed in variables of
the type IIB configurations, to pairs $(x^6,W_4)$ of $x^6$ positions
and Wilson lines along $4$ of the NS branes. We also have the torus
${\overline E}$, with $k k'/r$ points on it. Their positions
correspond, in type IIB language, to the pairs $(x^4,W_6)$ of $x^4$
positions and Wilson lines along 6 of the NS$'$ branes. Notice how the
complex structure parameters $\tau$ of the tori are equal because
Wilson lines around a circle of radius $R$ live on circles of radius
$1/R$.

The $\ZZ_{k/r}$ symmetry implies that the $k$ points on $E$ actually
are grouped as $k/r$ copies of a set of $r$ independent
points. Similarly, the $k k'/r$ points on ${\overline E}$ group in
$k/r$ copies of a set of $k'$ independent points. The $\ZZ_{k/r}$
action in the IIB configuration amounts to a simultaneous shift in
$x^4$ (by a fraction $r/k$ of the total period) and $x^6$ (by a
fraction $p/k$ of the total period), so it acts accordingly as a
simultaneous shift in $E$ {\em and} ${\overline E}$, with the correspondent
change of copies of marked points in the tori.

The quotient can be taken by choosing a representative for each
$\ZZ_{k/r}$ orbit. This simply means we can use $\ZZ_{k/r}$ to
restrict the range of the $x^4$ positions of NS$'$ branes to a
circle of radius a fraction $r/k$ of the total one, and to only one
copy of the points in ${\overline E}$. So we are considering a torus
${\overline E'}$ of complex structure parameter $(r/k) \tau$ (in the
type IIB construction, this corresponds to the area of the $k \times
k'$ box model being $k/r$ times smaller than the area of the $k \times
k k'/r$ box model), and $k'$ points on it.
Notice that the range of $x^6$ positions of points
in $E$ cannot be restricted because we have already used all of the
$\ZZ_{k/r}$ symmetry.

Since the radius of $x^4$ has been reduced, the Wilson lines around 4
live in a circle $k/r$ times larger, so we have to consider a new
torus $E'$, instead of the `old' $E$. The torus $E'$ also has complex
structure parameter $(r/k) \tau$. In this torus there are $r$
independent marked points, and for each we have $k/r$ copies at
symmetric positions.

So, the parameters in the theory are: the complex structure for the
tori $E'$, ${\overline E'}$, the choice of the $k'$ points in
${\overline E'}$, and the choice of $r$ points in $E'$ (since the
addition of the copies implies no choice at all). As before, since the
complex structure of the tori is the same, we can identify them as
long as we organize the points in two kinds. The space of parameters
is then analogous to those analyzed above, $\CM_{1,r,k'}$, and the
duality group is its fundamental group. In the example in section III,
we had $k=3$, $k'=1$, $p=1$, so only the overall \SL\ remains. This
agrees with the duality group proposed in \cite{Ibanez,ks} for this
model.

Finally, let us comment on the duality properties of the cylindrical
models. These can be obtained by starting with an appropriate elliptic
model whose duality group is known, and taking a certain limit in the
space of parameters. The duality group of the degenerated model is the
subgroup of the original duality group that commutes with this
limit.\footnote{This statement is familiar from several examples in
string duality.}

As a simple example, let us consider the degeneration of an elliptic
\ntwo\ $SU(N)^k$ model to a cylindrical $SU(N)^{k-1}$, introduced
above.  The space of parameters of the initial model is
$\CM_{1,k}$. It is easy to see that the limit merely involves the
degeneration of the M-theory torus $E$ to a cylinder; the $k$ points
remain at finite distance and their parameters remain finite in the
limit. The cylinder can be treated as a genus zero Riemann surface
with two ordered marked points. Thus, the space of parameters is the
moduli space of genus zero Riemann surfaces with $k+2$ marked points,
$k$ of which are unordered, and two of which are ordered
\cite{witMa}. The space is denoted $\CM_{1,k+2;2}$, and the duality
group is its fundamental group.

The \none\ examples can be studied analogously. The simplest
degeneration of a $k\times k'$ box elliptic model corresponds to
sending to infinity the areas of the boxes in one row or in one
column, while keeping the rest of the boxes of finite area, and
keeping the NS and NS$'$ branes at fixed positions and with fixed
Wilson lines. In the M-theory version, where the structure of the
space of parameters is most clear, this corresponds to an (identical)
degeneration in the complex structures of the tori $E$ and ${\overline
E}$. The space of parameters is the moduli space of a Riemann surface
of genus zero with $2$ ordered marked points, $k$ unordered marked
point of one kind, and $k'$ unordered marked points of another. The
duality group of the cylindrical model is the fundamental group of
this space.

In conclusion, the brane configurations discussed here provide a very
simple construction of a large family of finite \none\ models and a
geometric interpretation of (at least part of) their duality groups.
More investigation into the coupling constants is needed, as discussed
in Appendix A.  Our understanding of the duality groups of the
non-finite models is less certain, though it seems likely that the
overall \SL\ at least is present.

\

\

A.H. thanks D. L\"ust and A. Zaffaroni for useful discussions.
A.U. would like to thank L. Ibanez for useful conversations.  We thank
M. Gremm for bringing his related work to our attention prior to its
publication.  The work of A.H. and M.J.S.~was supported in part by
National Science Foundation grant NSF PHY-9513835; that of M.J.S. was
also supported by the W.M.~Keck Foundation.  The work of A.U.~was
supported by the Ram\'on Areces Foundation.

\

\centerline{{\bf APPENDIX A: More on coupling constants}}

\

Let us briefly comment on some issues that arise in the definition of the
gauge coupling and theta angles of the gauge factors in terms of brane
variables. Starting with the simplest case of \nfour\ theories, we have
seen that the gauge coupling is given by
\be
\tau \;=\; \int_{46} B_{RR} + i \frac{R_4 R_6}{g_s l_s^2}
\ee
There are several ways of showing this. An interesting one is to `measure'
$\tau$ by computing the action $S=-2\pi i \tau$ for an instanton of the
gauge theory. In the brane construction, the instanton corresponds to an
euclidean D string wrapping the $T^2$. The contribution to the real part
of the action comes from the area the D string is stretching across, while
the imaginary part arises from its coupling to the RR 2-form.

Let us turn to the \ntwo\ theories. In this case there are several
gauge factors, corresponding to the different regions $Q_i$ of the
two-torus bounded by the $(i-1)^{th}$ and the $i^{th}$ NS branes. It
is easy to see that a naive extension of the previous formula
(i.e. $\tau$ as the integral of the area and the two-form on the
relevant region) cannot be the complete answer. By performing a gauge
transformation on the 2-form, $B \to B+ d\lambda$, $\tau$ would change
by a boundary term $\oint_{\partial Q_i} \lambda$. The correct
expression can be obtained by computing the action of the euclidean D
string stretched across $Q_i$. It is actually simpler to argue in the
IIB S-dual version of the configuration, where a fundamental string
stretches between two D5 branes \footnote{When dealing with
configurations where the branes bend, this bending depends on the
value of the string coupling, and the effect of the S-duality on the
brane configuration would be complicated. In our case, however, the
branes are undergo no bending, and one can perform the duality without
changing the geometry of the configuration.}. The only difference with
the previous case is a contribution from the world-sheet boundary,
\be
\oint_{\partial Q_i} A_{m} \partial_{\sigma} X^m
\ee
where $X^m$ are the coordinates parametrizing the D5 brane,
$A_m$ is a 1-form living on it, and $\sigma$ parametrizes the world-sheet
boundary (in our case $\partial Q_i$). Since the boundary has two
disconnected components $l_i$, $l_{i+1}$ with different gauge fields
$A_i$, $A_{i+1}$, the complete expression for the theta angle is
\be
\theta_i = \int_{Q_i} B + \oint_{b} (A_{i} - A_{i-1})
\ee
where $b$ denotes the $x^4$ circle in the torus.
This is invariant under the gauge transformation of $B$ provided that the
gauge fields transform appropriately under it, $A\to A+\lambda$.
Thus in the \ntwo\ theories, the individual gauge couplings and theta
angles are controlled by the positions of the NS branes, and the Wilson
lines of their world-volume gauge fields around the compact dimension
$x^6$.

Let us finally turn to the \none\ theories. Each gauge factor arises from
a rectangle $R_{ij}$ bounded in $x^6$ by the $(i-1)^{th}$ and $i^{th}$ NS
branes and in $x^4$ by the $(j-1)^{th}$ and $j^{th}$ NS$'$ branes. These
branes carry gauge fields $A_{i-1}$, $A_{i}$ and $A'_{j-1}$, $A'_{j}$. The
theta angle has contributions from the rectangle and its boundary
\bel{thetaint}
\theta_{ij} = \int_{R_{ij}} B + \int_{a_i}(A'_{j-1} - A'_{j}) +
\int_{b_j} (A_{i}-A_{i-1})
\ee
where $a_i$ and $b_j$ denote the sides of the rectangle.
This is the simplified expression we introduced in section II. It is gauge
invariant under transformation of $B$ (with the one-forms compensating
the boundary term). However, it is not invariant under the gauge
transformations of the one-forms, since the range of line integral for
each gauge field is an interval with two points as boundary. The natural
solution is that there exist fields $\phi_{i,j}$ living at the
intersection of the NS and NS$'$, and which transform like axions under the
gauge transformations of the one-forms on the corresponding NS and NS$'$
branes. The complete expression for $\theta_{ij}$ then includes an
additional contribution $\phi_{i,j}$ $- \phi_{i,j-1}$ $-\phi_{i-1,j}$
$+\phi_{i-1,j-1}$.

There are natural candidates for these states, as strings stretching
between the $i^{th}$ NS and $j^{th}$ NS$'$ branes. Since there are two
directions in which the NS and NS$'$ branes can separate (8 and 9),
the states form a hypermultiplet under the \ntwo\ supersymmetry
unbroken by the NS fivebranes at the four dimensional
intersection. The presence of the D5 branes breaks supersymmetry
further, but only decomposes this \ntwo\ multiplet in \none\
multiplets, without projecting out any fields.  This is so because the
D5 branes do not forbid any brane motion for the NS fivebranes.

In the absence of the D5 branes, many interesting properties of these
fields -- for instance, the coupling of these fields to the six
dimensional fields living on the world-volume of the NS fivebranes --
could be analyzed using open string perturbation theory in the S-dual
version of the construction.  In particular, it should be possible to
study how gauge invariance in \eref{thetaint} is restored by the
contribution from the fields at the D5-D5$'$ intersections, as we have
proposed.  The contributions of these fields to the gauge coupling
constant could also be computed as the action of a fundamental string
stretched across the rectangle. However, it is not completely clear
whether in this last calculation the effects of the NS branes (S-dual
of the original D5 branes) can be ignored.

This type of computation may also help in understanding how the extra
marginal operators of the field theory may be represented in the brane
picture. We hope that future developments along these lines will allow for
a description of the complete parameter space and duality group in
\none\ theories.

\

\centerline{{\bf APPENDIX B:  Proof of Marginality}}

\

In this section we prove that a model with a grid laid out by NS and
NS$'$ branes, and with $n_{i,j}$ D5 branes stretched across the box in
the $({i,j})$ position, can have at least one marginal operator.  We
also prove the existence of additional marginal couplings when
$n_{i,j}=N$ for all $i,j$.

As discussed in section 2, each box has a gauge group
with a gauge coupling $g_{i,j}$, while each corner at
which four boxes intersect has two Yukawa couplings $h_{i,j}^+$,
$h_{i,j}^-$.
In previous sections we rescaled the holomorphic
version of these  Yukawa couplings to unity by redefining the fields.
Here it is useful to rescale them back into
the superpotential so that the techniques of \SEC{fieldtheory}
may be used without modification.  We define them by rewriting
\Eref{HVD}
\be
W=\sum_{i,j} h_{i,j}^+ H_{i,j}V_{i+1,j}D_{i+1,j+1}
-\sum_{i,j} h_{i,j}^- H_{i,j+1}V_{i,j}D_{i+1,j+1}.
\ee
We will show in this section that
\bel{toshow}
\sum_{i,j} A(g_{i,j}) \propto
\sum_{i,j} n_{i+1,j}A(h_{i,j}^+) + n_{i,j+1}A(h_{i,j}^-)
\ee
where $A(g), A(h)$ are defined in Eqs.~\eref{betah}-\eref{betaSV}.
As discussed in \SEC{fieldtheory}, this linear relation
allows the theory to have a marginal operator.

Since all of the Yukawa couplings are dimensionless,
this relation can only be true if the constant terms in
on the left hand side of \eref{toshow} cancel.
This sum is
\bel{bzero}
\sum_{i,j} (b_0)_{i,j} = \half\left(6n_{i,j}-
n_{i,j+1}-n_{i+1,j}-n_{i-1,j-1}-n_{i,j-1}-n_{i-1,j}-n_{i+1,j+1}\right)
\ee
which clearly vanishes.

It remains to show that the terms proportional to the anomalous
dimensions satisfy \eref{toshow}.  Begin with
\be \begin{array}{rcl}
\sum_{i,j} A(g_{i,j}) =&\sum_{i,j} \half\Big[
n_{i,j+1}\gamma(V_{i,j})+n_{i+1,j}\gamma(H_{i,j})
+n_{i-1,j-1}\gamma(D_{i,j})\\ \\
& \ \  +n_{i,j-1}\gamma(V_{i,j-1})
+n_{i-1,j}\gamma(H_{i-1,j})+n_{i+1,j+1}\gamma(D_{i+1,j+1})\Big]
\end{array}
\ee
where $\gamma(\phi)$ is the anomalous mass dimension of
the corresponding field (see \SEC{fieldtheory} for definitions.)
This can be resummed to read
\be \begin{array}{rcl}
\sum_{i,j} A(g_{i,j})&=\half\sum_{i,j}&
\gamma(H_{i,j})\left[n_{i,j}+n_{i+1,j}\right]+
\gamma(V_{i,j})\left[n_{i,j}+n_{i,j+1}\right]+
\gamma(D_{i+1,j+1})\left[n_{i,j}+n_{i+1,j+1}\right]\\ \\
&=\half\sum_{i,j}&
n_{i+1,j}\left[\gamma(V_{i,j})+\gamma(H_{i+1,j})
+\gamma(D_{i+1,j+1})\right]  \\ \\
& & + n_{i,j+1}\left[\gamma(H_{i,j})+\gamma(V_{i,j+1})
+\gamma(D_{i+1,j+1})\right]  \\ \\
& &+
\gamma(D_{i+1,j+1})\left[n_{i,j}+n_{i+1,j+1}-n_{i+1,j}-n_{i,j+1}\right]
\end{array}
\ee
Recognizing the first line of the final expression as $n_{i+1,j}
A(h_{i,j}^+)$ and the second as $n_{i,j+1} A(h_{i,j}^-)$, we find that
\Eref{toshow} is satisfied if \Eref{crosscheck} holds for all $(i,j)$.
As noted in \SEC{margops}, gauge anomalies are automatically cancelled,
and bending of each NS and NS$'$ brane is minimized.

Now consider the case when $n_{i,j}=N$.  Additional marginal couplings
follow from the results
\bel{moretoshow}\begin{array}{rcl}
\sum_{i=1}^k A(g_{i,j}) &=&
 N\sum_{i=1}^k A(h_{i,j}^+) + A(h_{i,j}^-) \\ \\
\sum_{j=1}^{k'} A(g_{i,j}) &=&
 N\sum_{j=1}^{k'} A(h_{i,j}^+) + A(h_{i,j}^-) \\ \\
\sum_{p=1}^P A(g_{i+p,j+p}) &=&
 N\sum_{p=1}^P A(h_{i+p,j+p}^+) + A(h_{i+p,j+p}^-) 
\end{array}
\ee
for all $i,j$, where $P$ is the smallest number with $k$ and $k'$ as a
common factor.  Since $i=i+P$ mod $k$ and $j=j+P$ mod $k'$, the boxes
$\{i+p,j+p\}$ form a continuous diagonal loop aroung the torus. Each
linearly independent relation gives a separate marginal operator; thus
we have one marginal coupling for each row, each column, and each
diagonal from lower left to upper right.  Note the sum of the first
line over $j$ and the sum of the second line over $i$ both equal
\Eref{toshow}; thus the rows and columns give $k-1$ and $k'-1$
independent relations respectively.  Similarly, the number of
independent relations stemming from the third line is one less than
$kk'/P$.

The first relation follows from the following.
\be \begin{array}{rll}
\sum_{i=1}^k &A(g_{i,j})  \\
&=\half N\sum_{i=1}^k &\Big[
\gamma(V_{i,j})+\gamma(H_{i,j})
+\gamma(D_{i,j})+\gamma(V_{i,j-1})
+\gamma(H_{i-1,j})+\gamma(D_{1,j+1})\Big] \\ \\
&=\half N\sum_{i=1}^k&\Bigg(\Big[
\gamma(V_{i+1,j})+\gamma(H_{i,j})
+\gamma(D_{i+1,j+1})\Big] \\
& & \ + \Big[\gamma(V_{i,j-1})
+\gamma(H_{i-1,j})+\gamma(D_{i,j})\Big] \Bigg)\\
&= N\sum_{i=1}^k &\left(A(h_{i,j}^+) + A(h_{i,j}^-)\right)
\end{array}
\ee
Analogous proofs establish the other relations.

   \bibliography{finite}        
\bibliographystyle{utphys}
\end{document}